\documentclass{ptephy_v1}

\preprintnumber{XXXX-XXXX} 

\usepackage{isotope}
\usepackage{amsmath}	
\usepackage{amssymb}    
\usepackage{booktabs}
\usepackage{upgreek}
\usepackage{xcolor}

\usepackage{caption}
\usepackage{subcaption}

\usepackage{graphics} 
\usepackage{url} 

\usepackage[separate-uncertainty=true]{siunitx} 
\newcommand{\MM}[1]{\textcolor{black}{#1}}

\begin{document}

\title{Application and modeling of an online distillation method to reduce krypton and argon in XENON1T}

\author[1]{E.~Aprile}
\author[2]{K.~Abe}
\author[3]{F.~Agostini}
\author[4]{S.~Ahmed Maouloud}
\author[5]{M.~Alfonsi}
\author[6]{L.~Althueser}
\author[7]{E.~Angelino}
\author[8]{J.~R.~Angevaare}
\author[9]{V.~C.~Antochi}
\author[10]{D.~Ant\'on Martin}
% ...
\affil[1]{Physics Department, Columbia University, New York, NY 10027, USA}
\affil[2]{Kamioka Observatory, Institute for Cosmic Ray Research, and Kavli Institute for the Physics and Mathematics of the Universe (WPI), University of Tokyo, Higashi-Mozumi, Kamioka, Hida, Gifu 506-1205, Japan}
\affil[3]{Department of Physics and Astronomy, University of Bologna and INFN-Bologna, 40126 Bologna, Italy}
\affil[4]{LPNHE, Sorbonne Universit\'{e}, Universit\'{e} de Paris, CNRS/IN2P3, 75005 Paris, France}
\affil[5]{Institut f\"ur Physik \& Exzellenzcluster PRISMA$^{+}$, Johannes Gutenberg-Universit\"at Mainz, 55099 Mainz, Germany}
\affil[6]{Institut f\"ur Kernphysik, Westf\"alische Wilhelms-Universit\"at M\"unster, 48149 M\"unster, Germany}
\affil[7]{INAF-Astrophysical Observatory of Torino, Department of Physics, University  of  Torino and  INFN-Torino,  10125  Torino,  Italy}
\affil[8]{Nikhef and the University of Amsterdam, Science Park, 1098XG Amsterdam, Netherlands}
\affil[9]{Oskar Klein Centre, Department of Physics, Stockholm University, AlbaNova, Stockholm SE-10691, Sweden}
\affil[10]{Department of Physics \& Kavli Institute for Cosmological Physics, University of Chicago, Chicago, IL 60637, USA}
% ...

\collaborator{XENON Collaboration\email{xenon@lngs.infn.it}\email{michael.murra@columbia.edu}}

\date{\today} 

\begin{abstract}%
A novel online distillation technique was developed for the XENON1T dark matter experiment to reduce intrinsic background components more volatile than xenon, such as krypton or argon, while the detector was operating. The method is based on a continuous purification of the gaseous volume of the detector system using the XENON1T cryogenic distillation column. A krypton-in-xenon concentration of \SI{360(60)}{ppq} was achieved. It is the lowest concentration measured in the fiducial volume of an operating dark matter detector to date. A model was developed and fit to the data to describe the krypton evolution in the liquid and gas volumes of the detector system for several operation modes over the time span of 550 days, including the commissioning and science runs of XENON1T. The online distillation was also successfully applied to remove \isotope[37]{Ar} after its injection for a low energy calibration in XENON1T. This makes the usage of \isotope[37]{Ar} as a regular calibration source possible in the future. The online distillation can be applied to next-generation LXe TPC experiments to remove krypton prior to, or during, any science run. The model developed here allows further optimization of the distillation strategy for future large scale detectors.
\end{abstract}

\subjectindex{xxxx, xxx}

\maketitle

\section{Introduction}
\label{sec:introduction}
Intrinsic radioactive noble gas contaminants such as \isotope[85]{Kr} and \isotope[222]{Rn} are the main contributors to the background in today's large scale liquid-xenon-based dark matter experiments \cite{Aprile:2017aty,Aprile_2020,AKERIB2020163047,PandaX-4T,Aalbers_2016}, as well as neutrinoless double beta decay experiments \cite{PhysRevC.97.065503,Alvarez:2012sma}. Their removal is of crucial importance for reaching the target sensitivities with growing demands on lowering backgrounds. A well-established technology for this removal is the application of cryogenic distillation columns that employ the differences in vapor pressure between the contaminant and xenon. More volatile components such as krypton are enriched in the gaseous xenon (GXe) phase, while less volatile constituents such as radon accumulate in the liquid xenon (LXe) \cite{Abe:2008py,Wang:2014,Aprile:2016xhi,Bruenner:2016ziq,Aprile:2017radon_xe100}.

The isotope \isotope[85]{Kr} is a $\upbeta$-emitter with an endpoint energy of \SI{687}{keV} and a half-life of \SI{10.76}{yr}. It is anthropogenically produced in uranium and plutonium fission and is released in the atmosphere by nuclear weapon tests and nuclear reprocessing plants. The abundance of \isotope[85]{Kr} in natural krypton is typically reported to be $\isotope[85]{Kr}/\isotope[\textrm{nat}]{Kr} \sim 10^{-11}$ \cite{Du2003}. Since xenon is extracted from air by fractional distillation, a small portion of natural krypton is contained within the xenon, typically at the level of ppm (\SI{e-6}{mol/mol}). Xenon with a lower krypton concentration (\isotope[\textrm{nat}]{Kr}/Xe  $\sim$\SI{10}{ppb} (\SI{e-9}{mol/mol})) can be purchased from industrial vendors. Current and future dark matter experiments require natural krypton-in-xenon concentrations at the \si{ppt} (\SI{e-12}{mol/mol}) level or below \cite{Aprile_2020}. Typically, it is assumed that the release of krypton from the detector components is negligible. Therefore, it needs to be removed from the xenon just once before the dark matter search. This removal is conventionally done by offline distillation or gas chromatography campaigns \MM{\cite{AKERIB201880,AKERIB2020163047}} before the start of an experiment, \MM{where both technologies are able to reach the required purity.}

In the case of the XENON1T experiment, the detector was initially filled with about \SI{3.2}{tonnes} of xenon without offline krypton removal. After the verification of the liquid xenon time projection chamber (LXe TPC) functionality, a novel online krypton distillation technique was developed using the existing XENON1T distillation column \cite{Aprile:2016xhi} to reduce the krypton-in-xenon concentration while the detector was operated.

The same online distillation technique was applied to remove the more volatile noble gas argon from xenon. The radioactive isotope \isotope[37]{Ar} was introduced into the XENON1T detector just before decommissioning for calibration purposes \cite{Aprile:Ar37}. Its decay via electron capture allowed the study of the detector response at low energies of \SI{2.8}{keV} for K-shell and \SI{0.27}{keV} for L-shell transitions \cite{TabRad_v7}. However, its half-life of \SI{35.01}{d} is too long for it to decay away in the scope of a dark matter search. A regular use is possible through the active removal of the residual argon after the calibration via the online distillation of volatile impurities.

The aforementioned new online distillation technique for argon and krypton is presented in this paper. In section \ref{sec:exp_setup}, the XENON1T detector system is summarized with a focus on the involved systems for the online distillation. In section \ref{sec:xe1t_kr_model}, a model is introduced in order to describe the concentration evolution of the more volatile noble gases in the gaseous and liquid xenon volumes of the detector for each operation mode. Section \ref{sec:model_fit} describes the fit of the model to the krypton data obtained from the event rate inside the LXe TPC itself, as well as from extracted xenon samples. Furthermore, the online removal of \isotope[37]{Ar} is presented in section \ref{sec:argon}. Section \ref{sec:conclusion} gives a conclusion and outline for possible future applications of the newly developed method.
\section{Experimental setup}
\label{sec:exp_setup}
The XENON1T experiment (decommissioned in December 2018) was located underground in the Laboratori Nazionali del Gran Sasso (LNGS), Italy, and utilized a total of \SI{3.2}{tonnes} of xenon. The LXe TPC inside the cryostat enclosed about \SI{2}{tonnes}, while the surrounding \SI{1.2}{tonnes} were employed as a passive shield. The cryostat was placed inside a \SI{10}{m} wide and \SI{10}{m} tall water tank equipped with an active Cherenkov muon veto system \cite{Aprile:2014zvw}, in order to shield against environmental radioactivity and remaining cosmic radiation. A service building next to the water tank hosted a cryogenic distillation column (DST), a purification (PUR) system, and a cryogenic (CRY) system, the relevant systems for this work. A complete overview about the different subsystems are summarized in Ref. \cite{Aprile:2017aty}.
\begin{figure}[!h]
\centering
\includegraphics[width=0.95\textwidth]{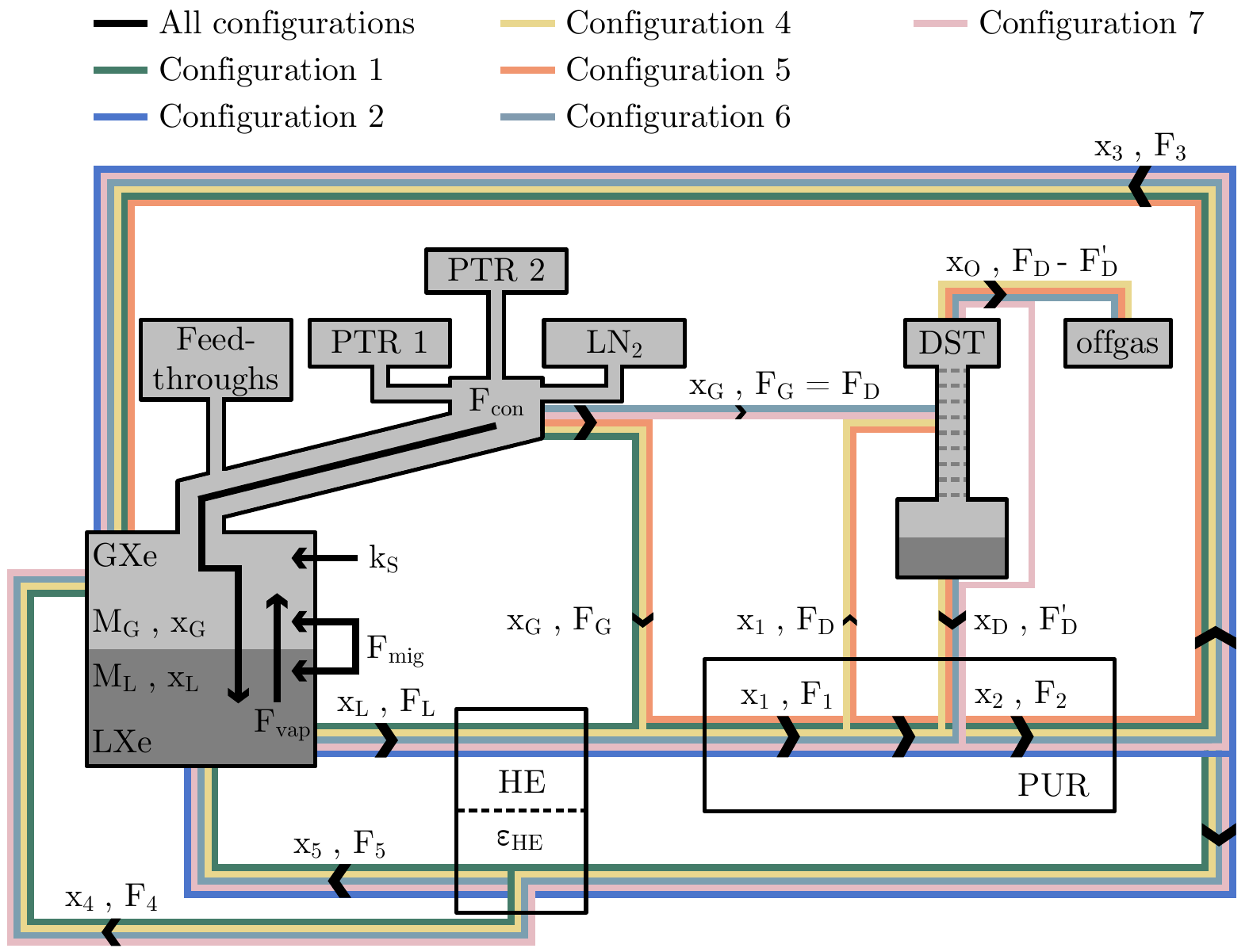}
\caption{Online distillation scheme: The liquid (gaseous) xenon phase LXe (GXe) in the cryostat contains the mass $M_{\mathrm{L}}$ ($M_{\mathrm{G}}$) and a solute concentration of $x_{\mathrm{L}}$ ($x_{\mathrm{G}}$). Different concentrations in different locations are denoted as $x_{\mathrm{i}}$ with their respective xenon mass flow $F_{\mathrm{i}}$. \MM{A constant solute particle flow $k_{\mathrm{S}}$ is entering the GXe volume to account for potential internal out-gassing or external leaks.} The global flow path depends on the detector configuration applied as indicated by the different colored lines. \MM{Details for each configuration are explained in section \ref{sec:xe1t_kr_model} and are summarized in table \ref{tab:configurations}. In Configuration 4 (gold), as example, xenon is extracted from the LXe volume with ($x_{\mathrm{L}}$, $F_{\mathrm{L}}$) and from the GXe with ($x_{\mathrm{G}}$, $F_{\mathrm{G}}$). The two flows mix at the PUR inlet into ($x_{\mathrm{1}}$, $F_{\mathrm{1}}$). A fraction ($x_{\mathrm{1}}$, $F_{\mathrm{D}}$) is distilled with the DST system and is returned with ($x_{\mathrm{D}}$, $F^{\prime}_{\mathrm{D}}$) mixing back into ($x_{\mathrm{2}}$, $F_{\mathrm{2}}$) with a lower solute concentration. From there, the flow is split and a fraction ($x_{\mathrm{3}}$, $F_{\mathrm{3}}$) returns directly into the GXe volume. The remaining flow goes into the HE, where the xenon is partially liquefied due to the limited HE efficiency $\epsilon_\mathrm{HE}$, and a flow ($x_{\mathrm{5}}$, $F_{\mathrm{5}}$) goes into the LXe volume. The flow ($x_{\mathrm{4}}$, $F_{\mathrm{4}}$) stays gaseous and returns to the GXe volume.} Note that in Configuration 3 only the black lines are relevant as no xenon is extracted for purification.}
\label{fig:basic_idea_online_kr}
\end{figure}

The CRY system consists of three independent condensation towers as depicted in figure \ref{fig:basic_idea_online_kr}, two equipped with redundant pulse tube refrigerators (PTRs) and one with liquid nitrogen (LN$_{2}$) cooling as backup. The system keeps the xenon temperature constant during data taking. A double-walled vacuum insulated tube (cryopipe) connects the CRY system to the cryostat to carry LXe to the cryostat after re-condensation. 
LXe is extracted and evaporated from the cryostat via a tube-in-tube heat exchanger inside the cryopipe and the GXe produced is further warmed up with the help of two parallel-plate heat exchangers installed in series. For simplicity, the series of heat exchangers is treated as one heat exchanger referred to as HE. The HE outlet is subsequently guided to the PUR system. Most of the purified GXe returns to the other side of this HE to liquefy the xenon again and feed it back into the detector. A small gas fraction is guided directly into a diving bell system (omitted in figure \ref{fig:basic_idea_online_kr}) to regulate the liquid level inside the LXe TPC. Additionally, for purification purposes, it is possible to extract a fraction of the evaporating xenon from the cryostat at three different locations of the CRY system (feedthroughs, cryopipe, condensers) and send it to either the DST or the PUR system before re-condensation.

\begin{figure}[!h]
\centering
\includegraphics[width=0.7\textwidth]{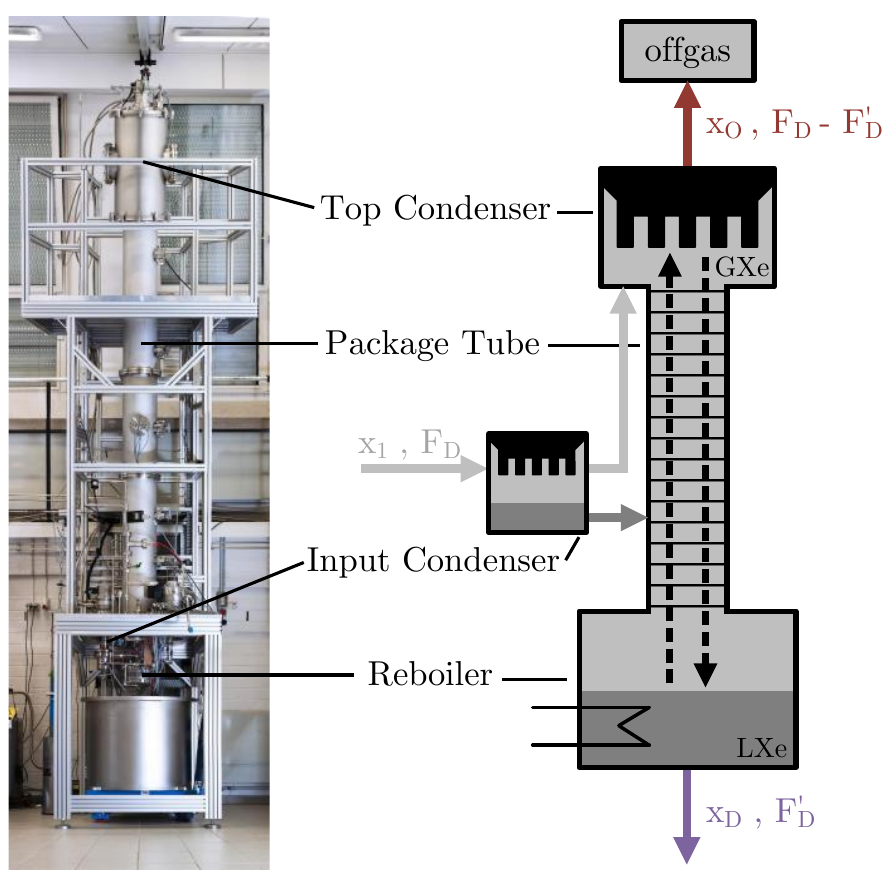}
\caption{(Left) Distillation column in the University of Münster before shipment to LNGS. (Right) Scheme of the distillation process showing the in-coming and out-going krypton concentrations $x_{\mathrm{i}}$ and xenon flows $F_{\mathrm{i}}$ as described in the text.}
\label{fig:dst_scheme}
\end{figure}

The PUR system continuously recirculates xenon extracted from the CRY system to remove electronegative impurities such as water and oxygen. These impurities can potentially suppress the light and charge signals in the LXe TPC. The PUR system is divided into two branches, each consisting of a high-purity pump, a flow controller, as well as a gas purifier. Furthermore, the PUR system acts as a xenon gas distributor between different subsystems. For a clearer visualization, the PUR system in figure \ref{fig:basic_idea_online_kr} shows only the distribution lines relevant for this work.
The DST system, designed to remove krypton from xenon, consists of four key components, namely an input condenser, a package tube, a reboiler and a top condenser. A scheme is shown in figure \ref{fig:dst_scheme} along with a picture of the set-up in the University of Münster before shipment to LNGS. The xenon enters the DST system with a krypton concentration $x_{\mathrm{1}}$ and flow $F_{\mathrm{D}}$ and is partially liquefied in the input condenser. From there, GXe and LXe are fed into the package tube at different heights. The reboiler at the bottom contains a liquid xenon volume that is partially evaporated, while the top condenser liquefies again the up-going xenon gas. In this manner, a counter-flow of up-going GXe and down-going LXe is established along the surface of the package tube, so that more volatile gases than xenon, such as argon or krypton, are enriched at the top and are depleted at the bottom. Here, ultra-pure xenon with a krypton concentration $x_{\mathrm{D}}$ can be extracted with a flow $F^{\prime}_{\mathrm{D}}$, typically about \SI{99}{\percent} of the feed flow $F_{\mathrm{D}}$. At the top, a small xenon fraction of \SI{1}{\percent} of $F_{\mathrm{D}}$ is extracted as krypton-enriched xenon offgas ($x_{\mathrm{O}}$, $F_{\mathrm{D}} - F^{\prime}_{\mathrm{D}}$) and is stored in bottles. The offgas is collected and distilled again in dedicated campaigns to minimize the xenon losses. 
\MM{The DST system's performance was determined by extracting samples during offline distillation campaigns, where xenon from bottles was distilled and filled into the XENON1T storage system at the process speed of \SI{72}{kg/d}. In one campaign, three bottles were measured with a commercial gas-chromatograph yielding an average concentration of \SI{453(53)}{ppb}. For a fourth bottle, a certificate from the delivering company stated a concentration of less than \SI{1000}{ppb} of krypton. The purified outlet sample was measured to have a concentration of \SI{730(140)}{ppq} measured from an extracted sample at the PUR system. Assuming a uniform probability for the unknown concentration of the fourth bottle between \num{0} and \SI{1000}{ppb}, the reduction factor is $(6.4^{+1.9}_{-1.4})\,\times\,10^{5}$ between feed and bottom product. Absolute concentrations $x_{\mathrm{D}}=$ \isotope[\mathrm{nat}]{Kr}/Xe\,\textless\,\SI{48}{ppq} (\SI{e-15}{mol/mol}) (\SI{90}{\percent} C.L.) were measured directly at the DST system's outlet when distilling xenon with input concentrations of about \SI{50}{ppb} during further offline distillation runs. All details are presented in Ref. \cite{Aprile:2016xhi}.}
%
%
%
%
%
%

For the online distillation method, the cryostat in combination with the CRY system is assumed to be a distillation-column-like system with an enrichment of more volatile noble gas species in the detector's GXe volume with respect to the LXe volume. In the following section, the online distillation method is explained for the case of krypton, although the method is applicable analogously to any noble gas species more volatile than xenon. The concept is based on the continuous purification of the GXe volume by the distillation column. This can be achieved by extracting the krypton-enriched xenon gas via the ports of the CRY system mentioned above. The krypton-free xenon from the DST system outlet returns to the PUR system and from there, back into the detector. This operation disturbs the krypton particle equilibrium between the LXe and GXe volumes, as the GXe volume now features a lower krypton concentration. As a consequence, krypton particles from the LXe volume migrate towards the GXe volume, from where they are removed again. Thus, a continuous krypton migration from LXe to GXe volume is established. In this way, the total xenon inventory of about \SI{3.2}{tonnes} can be purified by continuously processing about \SI{20}{kg} occupying the GXe volume.
\section{Online distillation model}
\label{sec:xe1t_kr_model}
In this section, an online distillation model is derived in order to describe the krypton concentration evolution in the GXe as well as in the LXe volume inside the detector over time. This paper covers a time span of 550 days from August 2016 until February 2018 including the commissioning phase, the first (SR0) and the second science run (SR1) of XENON1T \cite{SR0:SR1:onetonneyear:2018}. For the online distillation model, the in-going and out-going krypton flows in both volumes are taken into account, leading to a coupled differential equation system for a given detector configuration. The model can be analogously applied for the removal of any noble gas species more volatile than xenon, e.g. for the removal of argon. Therefore, the more general terminology \textit{solute concentration} is used in the following.

The PUR system continuously removes electronegative impurities from the LXe and GXe volumes during a background or a calibration run. The solute transport in the global system is derived based on this main operation. In the ideal case, the cryostat in figure \ref{fig:basic_idea_online_kr} can be simplified as a static LXe volume with mass $M_\mathrm{L}$ and solute concentration $x_\mathrm{L}$ with a GXe volume above with mass $M_\mathrm{G}$ and solute concentration $x_\mathrm{G}$. It is assumed that the solute in each volume is homogeneously distributed at all times. At equilibrium, the more volatile solute is enriched in the GXe. This enrichment can be described by the relative volatility $\alpha$, derived from Raoult's law \cite{McCabe:2005}, which is defined as the ratio between the vapor pressure of the noble gas solute $P_{\mathrm{S}}$ and that of xenon $P_{\mathrm{Xe}}$,
\begin{equation}
\label{eq:rel_volatility}
\alpha = \frac{P_{\mathrm{S}}}{P_{\mathrm{Xe}}}.
\end{equation}
From equation (\ref{eq:rel_volatility}) the concentration $x_{\mathrm{G}}$ in the GXe volume can be related to the concentration $x_{\mathrm{L}}$ in the LXe volume at equilibrium. For low solute concentrations ($\mathcal{O}(\mathrm{ppb})$ and below) the following relation holds true
\begin{equation}
\label{eq:equilibriumline_full}
x_{\mathrm{G}} =\frac{\alpha x_{\mathrm{L}}}{1+(\alpha-1)\ x_{\mathrm{L}}} \approx \alpha x_{\mathrm{L}},
\end{equation}
where it is assumed that $x_{\mathrm{L}}(\alpha-1)\ll 1$.

Away from equilibrium, solute particles can migrate from the liquid into the gas until equation (\ref{eq:equilibriumline_full}) is satisfied. This effect is introduced to the model via a migration term with a migration flow $F_{\mathrm{mig}}$ in units of a xenon mass flow; and is added in equations (\ref{eq:kr_online_model_gxe_standard_circulation}) and (\ref{eq:kr_online_model_lxe_standard_circulation}) describing the full model. More details on the migration term are given in Ref. \cite{Murra:2019:thesis}.

In reality, the system is not static as xenon evaporates with a mass flow $F_{\mathrm{vap}}$ in the cryostat due to the external heat input. This xenon needs to be re-condensed with a mass flow $F_{\mathrm{con}}$ with the help of one of the PTRs as shown in figure \ref{fig:basic_idea_online_kr}. It is possible that these two flows of xenon transport the solute between the two phases in the LXe TPC, but our data is not sensitive to the magnitude of this effect due to its degeneracy with $F_{\mathrm{mig}}$ and other free parameters of the model. In this model, evaporating xenon carries a solute concentration $\left(\alpha  x_{\mathrm{L}}\right)$, and condensing xenon at the PTR coldfinger carries a solute concentration $\left(\frac{x_{\mathrm{G}}}{\alpha}\right)$. It implies that the cryostat in combination with the CRY system acts as a distillation-column-like system with up to two distillation stages. This is further discussed below in subsection Configuration 3: No circulation.
Xenon from the GXe volume with a solute concentration $x_{\mathrm{G}}$ is extracted from the CRY system and guided to the PUR system with a mass flow $F_{\mathrm{G}}$. From the LXe volume, xenon with a solute concentration $x_{\mathrm{L}}$ is extracted and evaporated at a mass flow $F_{\mathrm{L}}$ through the HE. It is assumed that the solute concentration in the extracted xenon remains constant, as this is a pressure-driven flow, \MM{and that the solute concentration in general is independent from pressure changes in the flow path.}
Both xenon streams from the GXe and LXe volumes mix at the PUR system's inlet. The concentration $x_1$ in the summed mass flow $F_{\mathrm{1}} = F_{\mathrm{G}} + F_{\mathrm{L}}$ can be written as
\begin{equation}
\label{eq:standard_x1_conc}
x_1 = \frac{F_{\mathrm{G}}}{F_{1}}  x_{\mathrm{G}} + \frac{F_{\mathrm{L}}}{F_{1}}  x_{\mathrm{L}}. 
\end{equation}

At the PUR outlet, the solute particle flow $\left(x_2  F_{\mathrm{2}}\right)$ splits again. A solute particle flow $\left(x_3  F_{\mathrm{3}}\right)$ returns into the GXe volume, while a small fraction of this flow is guided directly into the bell to stabilize the LXe TPC liquid level. In this model, the gaseous bell volume is not separated from the rest of the GXe volume, as both are assumed to be in good contact allowing for fast mixing. Thus, the bell is omitted in figure \ref{fig:basic_idea_online_kr}. The remaining solute flow $\left(x_2  (F_{\mathrm{2}} - F_{\mathrm{3}})\right)$ flows into the HE for liquefaction. Due to its limited efficiency of $\epsilon_{\mathrm{HE}} < 1$, the in-going flow is divided into a solute particle flow $\left(x_4  F_{\mathrm{4}}\right)$ guided into the GXe and a solute particle flow $\left(x_5  F_{\mathrm{5}}\right)$ added to the LXe volume.
The solute concentration $x_4$ should be larger than $x_5$, scaling with an enrichment factor $\alpha_{\mathrm{HE}}$ of the size of the relative volatility $\alpha$ or even larger: $x_4 = \alpha_{\mathrm{HE}} x_5$. The respective xenon mass flows can be calculated to be $F_{\mathrm{5}} = \epsilon_{\mathrm{HE}}  (F_{\mathrm{2}} - F_{\mathrm{3}})$ and $F_{\mathrm{4}} = \left( 1 - \epsilon_{\mathrm{HE}}\right)  (F_{\mathrm{2}} - F_{\mathrm{3}})$.
Hence, the two solute particle flows to the GXe and LXe volumes can be written down as
\begin{alignat}{2}
\label{eq:kr_model_he_x4f4}
x_4  F_{\mathrm{4}} &= \frac{\alpha_{\mathrm{HE}}  x_2}{\epsilon_{\mathrm{HE}} + \alpha_{\mathrm{HE}} \left(1 - \epsilon_{\mathrm{HE}} \right)}  \left(1 - \epsilon_{\mathrm{HE}} \right)  (F_{\mathrm{2}} - F_{\mathrm{3}}), \\
\label{eq:kr_model_he_x5f5}
x_5  F_{\mathrm{5}} &= \frac{x_2}{\epsilon_{\mathrm{HE}} + \alpha_{\mathrm{HE}}  \left(1 - \epsilon_{\mathrm{HE}} \right)}  \epsilon_{\mathrm{HE}} (F_{\mathrm{2}} - F_{\mathrm{3}}).
\end{alignat}
The solute potentially enters the closed system either due to microscopic leaks in the global system or out-gassing from materials. Since the two phenomena cannot be distinguished, a single source parameter is introduced to the model as a constant solute particle flow $k_{\mathrm{S}}$ entering the GXe volume as indicated in figure \ref{fig:basic_idea_online_kr}.

Combining all effects, the solute concentration change over time $dx_{\mathrm{G}}/dt$ in the GXe and $dx_{\mathrm{L}}/dt$ in the LXe volumes can be described by the following set of differential equations:
%
%
%
%
%
%
%
%
%
%
\begin{equation}
\label{eq:kr_online_model_gxe_standard_circulation}
\begin{split}
M_{\mathrm{G}}\frac{dx_{\mathrm{G}}}{dt} =
&\ \overbrace{+\left[\alpha x_{\mathrm{L}} - x_{\mathrm{G}}\right] F_{\mathrm{mig}} }^{\substack{\text{(I)} \\ \text{migration}}} 
\overbrace{- \frac{x_{\mathrm{G}}}{\alpha} F_{\mathrm{con}}}^{\substack{\text{(II)} \\ \text{condensation}}} 
\overbrace{+ \alpha x_{\mathrm{L}} F_{\mathrm{vap}}}^{\substack{\text{(III)} \\ \text{evaporation}}} 
\overbrace{ - x_{\mathrm{G}}  F_{\mathrm{G}}}^{\substack{\text{(IV)} \\ \text{extraction}}}  \\
&\ \overbrace{+ x_{\mathrm{3}} F_{\mathrm{3}}}^{\substack{\text{(V)} \\ \text{return}}} 
\overbrace{+ x_{\mathrm{4}} F_{\mathrm{4}}}^{\substack{\text{(VI)} \\ \text{return HE}}} 
\overbrace{+ k_{\mathrm{S}}}^{\substack{\text{(VII)} \\ \text{source}}}, 
\end{split}
\end{equation}
%
%
%
\begin{equation}
\label{eq:kr_online_model_lxe_standard_circulation}
\begin{split}
M_{\mathrm{L}}\frac{dx_{\mathrm{L}}}{dt} =
&\ \overbrace{-\left[\alpha x_{\mathrm{L}} - x_{\mathrm{G}}\right] F_{\mathrm{mig}} }^{\substack{\text{(I)} \\ \text{migration}}} 
\overbrace{+ \frac{x_{\mathrm{G}}}{\alpha} F_{\mathrm{con}}}^{\substack{\text{(II)} \\ \text{condensation}}} 
\overbrace{- \alpha x_{\mathrm{L}} F_{\mathrm{vap}}}^{\substack{\text{(III)} \\ \text{evaporation}}} 
\overbrace{ - x_{\mathrm{L}}  F_{\mathrm{L}}}^{\substack{\text{(IV)} \\ \text{extraction}}}  \\
&\ \overbrace{+ x_{\mathrm{5}} F_{\mathrm{5}}}^{\substack{\text{(V)} \\ \text{return HE}}}.
\end{split}
\end{equation}
In both equations, the term (I) corresponds to the migration (gas-phase enhancement at equilibrium), the term (II) to the condensation and the term (III) to the evaporation. For the GXe volume, the additional terms are the gas extraction (IV), the directly returning gas (V), the additional returning gas from the HE (VI) and the constant source term (VII). For the LXe volume, further terms are the extraction (IV), and the liquid returning from the HE (V). Each solute particle flow is divided by the respective mass of the volume to model a change in the solute concentration rather than in the number of solute particles. Furthermore, the sign of each term indicates whether the solute is leaving ($-$) or entering ($+$) a volume.

One additional remark is that the LXe TPC measures the decay of \isotope[85]{Kr} particles, while the other methods, presented in section \ref{sec:model_fit}, determine the \isotope[\mathrm{nat}]{Kr} content within the samples. With \SI{10.76}{yr}, the half-life of \isotope[85]{Kr} is much longer than the time period investigated here. Thus, a krypton removal term due to its decay can be neglected for the differential equations.

During the time period that this paper concerns, the GXe circulation loop of XENON1T, comprising the PUR and DST subsystems, was operated in seven distinct configurations. The set of differential equations describing the solute transport must be tailored for a given configuration. For each configuration, table \ref{tab:configurations} shows the terms in equations (\ref{eq:kr_online_model_gxe_standard_circulation}) and (\ref{eq:kr_online_model_lxe_standard_circulation}) that are included, and figure \ref{fig:basic_idea_online_kr} shows the corresponding flow path. In the following, each configuration is briefly explained.
\begin{table}[!ht]
\caption{Description of the detector configurations. The terms (I), (II), and (III) in equations (\ref{eq:kr_online_model_gxe_standard_circulation}) and (\ref{eq:kr_online_model_lxe_standard_circulation}) are present for all configurations, and thus omitted here.}
\label{tab:configurations}
\centering
\begin{tabular}{c|l|cccc|cc}
Name & Description                                                                                                        & \multicolumn{4}{c|}{$dx_{\mathrm{G}}/dt$}                       & \multicolumn{2}{c}{$dx_{\mathrm{L}}/dt$} \\
     &                                                                                                                    & IV         & V          & VI         & VII        & IV           & V           \\ \hline
C1 & \begin{tabular}[l]{@{}l@{}}Standard purification\\ without distillation\end{tabular}                               & \checkmark & \checkmark & \checkmark & \checkmark & \checkmark   & \checkmark  \\ \hline
C2 & \begin{tabular}[l]{@{}l@{}}Evaporated-liquid-only\\ purification without distillation\end{tabular}                 &  $\times$ & \checkmark &  $\times$ & \checkmark & \checkmark   & \checkmark  \\ \hline
C3 & No circulation                                                                                                     & $\times$ & $\times$ & $\times$ & $\times$ & $\times$   & $\times$  \\ \hline
C4 & \begin{tabular}[l]{@{}l@{}}Standard purification\\ with distillation\end{tabular}                                  & \checkmark & \checkmark & \checkmark & \checkmark & \checkmark   & \checkmark  \\ \hline
C5  & \begin{tabular}[l]{@{}l@{}}Purification and distillation\\ of gas volume alone\end{tabular}                        & \checkmark & \checkmark &  $\times$ & \checkmark &  $\times$   &  $\times$  \\ \hline
C6  & \begin{tabular}[l]{@{}l@{}}Standard purification\\ with upgraded\\ gas-volume-only distillation\end{tabular}       & \checkmark & \checkmark & \checkmark & \checkmark & \checkmark   & \checkmark  \\ \hline
C7  & \begin{tabular}[l]{@{}l@{}}Standard purification with\\ upgraded gas-volume-only\\ radon distillation\end{tabular} & \checkmark & \checkmark & \checkmark & \checkmark & \checkmark   & \checkmark  \\ \hline
\end{tabular}
\end{table}

\subsection*{\textbf{Configuration 1: Standard purification without distillation}}
Some simplifications can be  made for this configuration: The two branches of the PUR system are equipped with heated getters. While electronegative impurities are efficiently removed, noble gases pass through these getters unaffected. Therefore, the inlet and outlet of the PUR system feature the same flow and solute concentration: $F_{\mathrm{2}} = F_{\mathrm{1}}$ and $x_{\mathrm{2}} = x_{\mathrm{1}}$, implying that $x_3 = x_1$. Additionally, the directly returning flow to the GXe volume is $F_3 = F_{\mathrm{G}}$ by design of the system. This leads to $(F_{\mathrm{2}} - F_{\mathrm{3}}) = F_{\mathrm{L}}$ for the flow to the HE. Due to its limited efficiency, more gas goes back to the GXe volume than being extracted. In order to keep the masses in the LXe and GXe volume constant, the condensation flow $F_{\mathrm{con}}$ needs to be larger than the evaporated flow $F_{\mathrm{vap}}$ in this configuration, giving
\begin{equation}
\label{eq:xe1t_kr_model_stand_circ_fvap}
F_{\mathrm{con}} = F_{\mathrm{vap}} + \left( 1 - \epsilon_{\mathrm{HE}} \right)  F_{\mathrm{L}}.
\end{equation}
\subsection*{\textbf{Configuration 2: Evaporated-liquid-only purification without distillation}}
In this configuration, xenon is extracted and subsequently evaporated only from the LXe volume with a concentration $x_{\mathrm{L}}$ at a flow $F_{\mathrm{L}}$, and thus $F_{\mathrm{G}} = 0$. This allows the extraction of xenon samples at the PUR system with the concentration $x_{\mathrm{L}}$ and gives direct insight to the krypton inside the LXe volume.

In the PUR system the mass flows are given by $F_{\mathrm{2}} = F_{\mathrm{1}} = F_{\mathrm{L}} $. At the PUR outlet, the flow back to the GXe volume is $F_{\mathrm{3}}$, where in this case it is equal to the flow going directly into the bell, and therefore lower than in other configurations. For the different concentrations $x_{\mathrm{3}} = x_{\mathrm{2}} = x_{\mathrm{1}} = x_{\mathrm{L}}$ holds true. The remaining flow of $\left(F_{\mathrm{L}} - F_{\mathrm{3}} \right)$ enters the HE. As a flow $F_{\mathrm{L}}$ is extracted through the HE from the LXe volume, the HE can liquefy a returning xenon flow equal to $\epsilon_{\mathrm{HE}}  F_{\mathrm{L}}$. It implies that
\begin{equation}
\left(F_{\mathrm{L}} - F_{\mathrm{3}} \right) \lesssim \epsilon_{\mathrm{HE}}  F_{\mathrm{L}}.
\end{equation}
Therefore, the assumption is made that the complete flow can be liquefied and returns into the LXe volume,%
\begin{alignat}{2}
x_4  F_4 &= 0, \\
x_5  F_5 &= x_{\mathrm{L}}  \left( F_{\mathrm{L}} - F_{\mathrm{3}} \right).
\end{alignat}
Based on that, the condensation flow in the CRY system needs to be 
\begin{equation}
F_{\mathrm{con}} = F_{\mathrm{vap}} + F_{\mathrm{3}}.
\end{equation}
The coupled differential equation system for this mode can be achieved by inserting above information into equations (\ref{eq:kr_online_model_gxe_standard_circulation}) and (\ref{eq:kr_online_model_lxe_standard_circulation}). For the GXe volume, the terms for gas extraction (IV) as well as for the additional gas return from HE (VI) are not present.
\subsection*{\textbf{Configuration 3: No circulation}}
In this configuration, e.g. during maintenance work on the PUR system, no xenon leaves or enters the detector system ($ F_{\mathrm{G}} =  F_{\mathrm{L}} = 0$). Thus, equations (\ref{eq:kr_online_model_gxe_standard_circulation}) and (\ref{eq:kr_online_model_lxe_standard_circulation}) are reduced to the migration (I), condensation (II) and evaporation (III) terms. The solute distribution between the GXe and LXe volume for the equilibrium case ($\frac{dx_{\mathrm{G}}}{dt} = \frac{dx_{\mathrm{L}}}{dt} = 0$) can be investigated. By neglecting the source (VII) term, and with equal evaporation and condensation flows ($F_{\mathrm{vap}} = F_{\mathrm{con}}$) based on equation (\ref{eq:kr_online_model_gxe_standard_circulation}) it follows that
\begin{equation}
x_{\mathrm{G}} = \alpha  \frac{\left( F_{\mathrm{mig}} + F_{\mathrm{con}} \right)}{\left( F_{\mathrm{mig}} +  \frac{F_{\mathrm{con}}}{\alpha} \right)}  x_{\mathrm{L}}.
\end{equation}
The only unknown parameter is the migration flow $F_{\mathrm{mig}}$. Therefore, the two extreme cases for a fast and a slow migration can be considered, that is
\begin{alignat}{2}
\label{eq:C3_lowfactor}
x_{\mathrm{G}} &= \alpha x_{\mathrm{L}} \quad \text{for} \quad F_{\mathrm{mig}} \gg F_{\mathrm{con}},\\
\label{eq:C3_highfactor}
x_{\mathrm{G}} &= \alpha^{2} x_{\mathrm{L}} \quad \text{for} \quad F_{\mathrm{mig}} \ll F_{\mathrm{con}}.
\end{alignat}
Due to the active cooling, the solute concentration in the GXe volume is enhanced by a factor between $\alpha$ (single-stage distillation) and $\alpha^2$ (two-stage distillation).
\subsection*{\textbf{Configuration 4: Standard purification with distillation}}
The proof of concept for the online distillation method was verified 
during the commissioning of XENON1T. The campaign lasted from 11 Aug to 22 Aug 2016, in parallel with the LXe TPC commissioning and without interference with other subsystems. Several xenon samples with the concentration $x_1$ were taken from the PUR system to monitor the krypton-in-xenon evolution.

The detector was operated in the standard purification mode to further decrease electronegative impurities, with the difference that a xenon mixture with solute flow $\left(x_1  F_{\mathrm{D}}\right)$ was extracted from the PUR system and guided to the DST system. Here, the xenon was purified from the solute with an offgas loss $\left(F_{\mathrm{D}} - F^{\prime}_{\mathrm{D}}\right)$. The solute particle flow returning to the PUR system is given by $\left(x_{\mathrm{D}}  F^{\prime}_{\mathrm{D}}\right)$.

As a consequence, the solute concentration at the PUR outlet is lower than in \textit{C1 (Standard purification without distillation)},
\begin{equation}
x_2 = \left(\frac{F_{\mathrm{1}} - F_{\mathrm{D}}}{F_{2}}\right)  x_1 + \frac{F^{\prime}_{\mathrm{D}}}{F_{2}}  x_{\mathrm{D}}.
\end{equation}
An assumption is made to further simplify the equation above: The offgas flow $\left(F_{\mathrm{D}} - F^{\prime}_{\mathrm{D}}\right)$ is neglected such that $F_{\mathrm{D}} = F^{\prime}_{\mathrm{D}}$, implying also $F_2 = F_1$. It follows with equation (\ref{eq:standard_x1_conc}) that
\begin{equation}
x_2 = \left(1-\frac{F_{\mathrm{D}}}{F_{1}}\right)  \left( \frac{F_{\mathrm{G}}}{F_{1}}  x_{\mathrm{G}} + \frac{F_{\mathrm{L}}}{F_{1}}  x_{\mathrm{L}}  \right) + \frac{F_{\mathrm{D}}}{F_1}  x_{\mathrm{D}}.
\end{equation}
The coupled differential equation system for \textit{C4 (Standard purification with distillation)} includes the same terms as \textit{C1 (Standard purification without distillation)}. The difference is that the PUR outlet mass flow $F_2$ now contains a reduced solute concentration $x_2$. Thus, all flows returning to the GXe and LXe volumes are also characterized by a lower solute concentration.
\subsection*{\textbf{Configuration 5: Purification and distillation of gas volume alone}}
In this configuration, xenon was solely extracted from the GXe volume to test the solute removal for a decreased exchange time with respect to \textit{C4 (Standard purification with distillation)} of this volume. A first attempt of this mode on 22 Aug 2016 was stopped after a few hours due to a broken circulation pump. After the pump replacement, the main campaign was performed from 24 Aug to 02 Sep 2016.

Since gas was only extracted from the GXe volume with particle flow $\left(x_{\mathrm{G}}  F_{\mathrm{G}}\right)$, the flow from the LXe volume was $F_{\mathrm{L}} = 0$. Thus, $F_{\mathrm{1}} = F_{\mathrm{G}}$ and $x_{\mathrm{1}} = x_{\mathrm{G}}$. Therefore, the samples taken from the PUR system give direct access to $x_{\mathrm{G}}$.
For the distillation, the same flow path as in \textit{C4 (Standard purification with distillation)} was used. The concentration in the PUR outlet is given by
\begin{equation}
\label{eq:kr_online_model_x2_gas_only_initial}
x_2 = \frac{F_{\mathrm{G}} - F_{\mathrm{D}}}{F_{2}}  x_1 + \frac{F^{\prime}_{\mathrm{D}}}{F_{2}}  x_{\mathrm{D}}.
\end{equation}
For a negligible offgas flow $\left( F_{\mathrm{D}} - F^{\prime}_{\mathrm{D}} \right)$, it follows that $F_{\mathrm{D}} = F^{\prime}_{\mathrm{D}}$, so that $F_{\mathrm{2}} = F_{\mathrm{1}} = F_{\mathrm{G}}$:
\begin{equation}
x_2 = \left(1 - \frac{F_{\mathrm{D}}}{F_{\mathrm{G}}} \right)  x_{\mathrm{G}} + \frac{F_{\mathrm{D}}}{F_{\mathrm{G}}}  x_{\mathrm{D}}.
\end{equation}
The HE is not operational in this mode as no liquid goes through it. Consequently, the returning flow from the PUR system goes fully back into the GXe volume with $F_{\mathrm{3}} = F_{\mathrm{G}}$, from which it follows that $F_{\mathrm{con}} = F_{\mathrm{vap}}$.

Inserting the information above into equations (\ref{eq:kr_online_model_gxe_standard_circulation}) and (\ref{eq:kr_online_model_lxe_standard_circulation}), one finds that the solute concentration change in the LXe volume depends only on the migration (I), condensation (II) and evaporation (III) terms, and no longer on the terms for extraction (IV) and return from HE (V). Furthermore, both equations are independent from the HE efficiency. This is different from the other configurations and makes this configuration more sensitive to $F_{\mathrm{mig}}$, $F_{\mathrm{con}}$ and $F_{\mathrm{vap}}$.
\subsection*{\textbf{Configuration 6: Standard purification with upgraded gas-volume-only distillation}}
The ultimate configuration required hardware modifications to combine \textit{C4 (Standard purification with distillation)} and \textit{C5 (Purification and distillation of gas volume alone)} with the advantage of purifying both volumes from electronegative impurities, while only distilling the GXe volume as rapidly as possible. For that, a direct connection between the GXe volume and the inlet of the DST system was installed as represented in figure \ref{fig:basic_idea_online_kr}.

Two short online distillation campaigns, from 28 Sep to 29 Sep 2016 and 13 Oct to 14 Oct 2016, were performed to verify the functionality of the final configuration.
Finally, a long-term online distillation campaign was performed from 28 Oct to 12 Dec 2016. During this operation, the total detector inventory was lowered by about \SI{6}{kg} per week due to the offgas flow. As a consequence, the liquid level inside the LXe TPC decreased by \SI{0.1}{mm} per week. However, the level was manually adjusted once per week to keep the impact on the LXe TPC performance negligible. Other influences on the detector operation were not observed.

In this configuration, the flow $F_{\mathrm{D}} = F_{\mathrm{G}}$ from the GXe volume is guided directly into the DST system and it contains the solute concentration $x_{\mathrm{G}}$. Note that no flow from the GXe volume is going directly to the PUR inlet. In parallel, only xenon from the LXe volume flows to the inlet of the PUR system with a solute content of $x_{\mathrm{1}} = x_{\mathrm{L}}$ at a flow $F_{\mathrm{1}} = F_{\mathrm{L}}$. The purified xenon returns from the DST to the PUR system as shown in figure \ref{fig:basic_idea_online_kr} with a solute concentration $x_{\mathrm{D}}$ and a flow $F^{\prime}_{\mathrm{D}}$. Under the assumption of a negligible offgas flow $\left( F_{\mathrm{D}} - F^{\prime}_{\mathrm{D}} \right)$ it follows that $F^{\prime}_{\mathrm{D}} = F_{\mathrm{D}} = F_{\mathrm{G}}$. Thus, the total flow at the PUR outlet is $F_{\mathrm{2}} = F_{\mathrm{G}} + F_{\mathrm{L}}$ and the concentration at this location can be calculated to be
\begin{equation}
x_2 =  \frac{F_{\mathrm{L}}}{F_{\mathrm{2}}}  x_{\mathrm{L}}   +  \frac{F_{\mathrm{G}}}{F_{\mathrm{2}}}  x_{\mathrm{D}}.
\end{equation}
After the PUR system outlet, the flow follows the path described in \textit{C1 (Standard purification without distillation)}.
\subsection*{\textbf{Configuration 7: Standard purification with upgraded gas-volume-only radon distillation}}
In addition to the online krypton distillation, two online radon distillation campaigns, from 19 Dec 2016 to 26 Jan 2017 and from 31 Jan to 02 Feb 2017, were performed to reduce the radon-induced background. For this operation, the same flow paths as for \textit{C6 (Standard purification with upgraded gas-volume-only distillation)} were used with the following difference: Radon as the less volatile noble gas accumulates at the bottom of the distillation column until it decays. The radon-depleted xenon exits the top of the column and returns to the PUR system as visualized in figure \ref{fig:basic_idea_online_kr}. The more volatile solutes enriched at the top are therefore not influenced and can pass through the DST system unaffected. Thus, the DST system's outlet flow contains the same solute concentration $x_{\mathrm{G}}$ as the inlet. It follows that
\begin{equation}
x_2 =  \frac{F_{\mathrm{L}}}{F_{\mathrm{2}}}  x_{\mathrm{L}}   +  \frac{F_{\mathrm{G}}}{F_{\mathrm{2}}}  x_{\mathrm{G}}.
\end{equation}
More details on the online radon distillation are given in Ref. \cite{Murra:2019:thesis}.

\section{Krypton removal}
\label{sec:model_fit}
The krypton concentration in the LXe and GXe volumes was monitored from August 2016 to February 2018 in order to observe the efficiency of the online distillation method for the different configurations as well as its evolution during the science runs. This was achieved using three different measurement methods: An on-site residual gas analyzer (RGA) system behind a LN$_2$-cooled coldtrap \cite{Brown:2013rga,Fieguth:2014}, an off-site rare gas mass spectrometer (RGMS) \cite{Lindemann:2013rgms}, and the electronic recoil (ER) event rate (ER rate) inside the LXe TPC itself.

The RGA system was utilized during \textit{C4 (Standard purification with distillation)} and \textit{C5 (Purification and distillation of gas volume alone)}, for a quick and direct feedback of the krypton decrease. The samples taken during \textit{C4 (Standard purification with distillation)} contained the concentration $x_\mathrm{1}$ and were a mixture of xenon extracted and evaporated from the LXe volume and xenon extracted from the GXe volume (referred to as $\hat{x}_\mathrm{1,RGA}$).

The samples during \textit{C5 (Purification and distillation of gas volume alone)} were extracted solely from the GXe volume (referred to as $\hat{x}_\mathrm{G,RGA}$). During \textit{C6 (Standard purification with upgraded gas-volume-only distillation)}, the krypton concentrations were below the RGA sensitivity and thus, no samples were measured with the RGA in this configuration.

The RGMS is capable of detecting trace amounts of natural krypton-in-xenon down to the ppq level \cite{Lindemann:2013rgms}. It allowed for the determination of the krypton concentration $x_\mathrm{L}$ within the LXe volume throughout the full time period investigated, including commissioning, science runs SR0 \cite{Aprile:2017iyp} and SR1 \cite{SR0:SR1:onetonneyear:2018}. For the extraction of the samples (referred to as $\hat{x}_\mathrm{L,RGMS}$), the detector operation was switched to \textit{C2 (Evaporated-liquid-only purification without distillation)}, where xenon was extracted and evaporated solely from the LXe volume. Additionally, on 25 May 2017, one sample with concentration $x_\mathrm{1}$ was extracted during \textit{C1 (Standard purification without distillation)} (referred to as $\hat{x}_\mathrm{1,RGMS}$).

The ER event rate (referred to as $\hat{x}_\mathrm{L,ER}$) data from the LXe TPC gave the most precise insight into the krypton evolution in the LXe volume as the krypton beta decay rate is proportional to the number of krypton atoms. Before the online distillation, the krypton concentration was on the order of ppb such that the overall detector event rate at energies up to \SI{200}{keV} was dominated by krypton beta events. Several selection criteria were applied to obtain the rate in a core volume of about \SI{725}{kg}. Days with low statistics or unstable detector conditions were also neglected. Further details are presented in Ref. \cite{Murra:2019:thesis}.

\begin{figure}[!ht]
\centering
\includegraphics[width=0.95\textwidth]{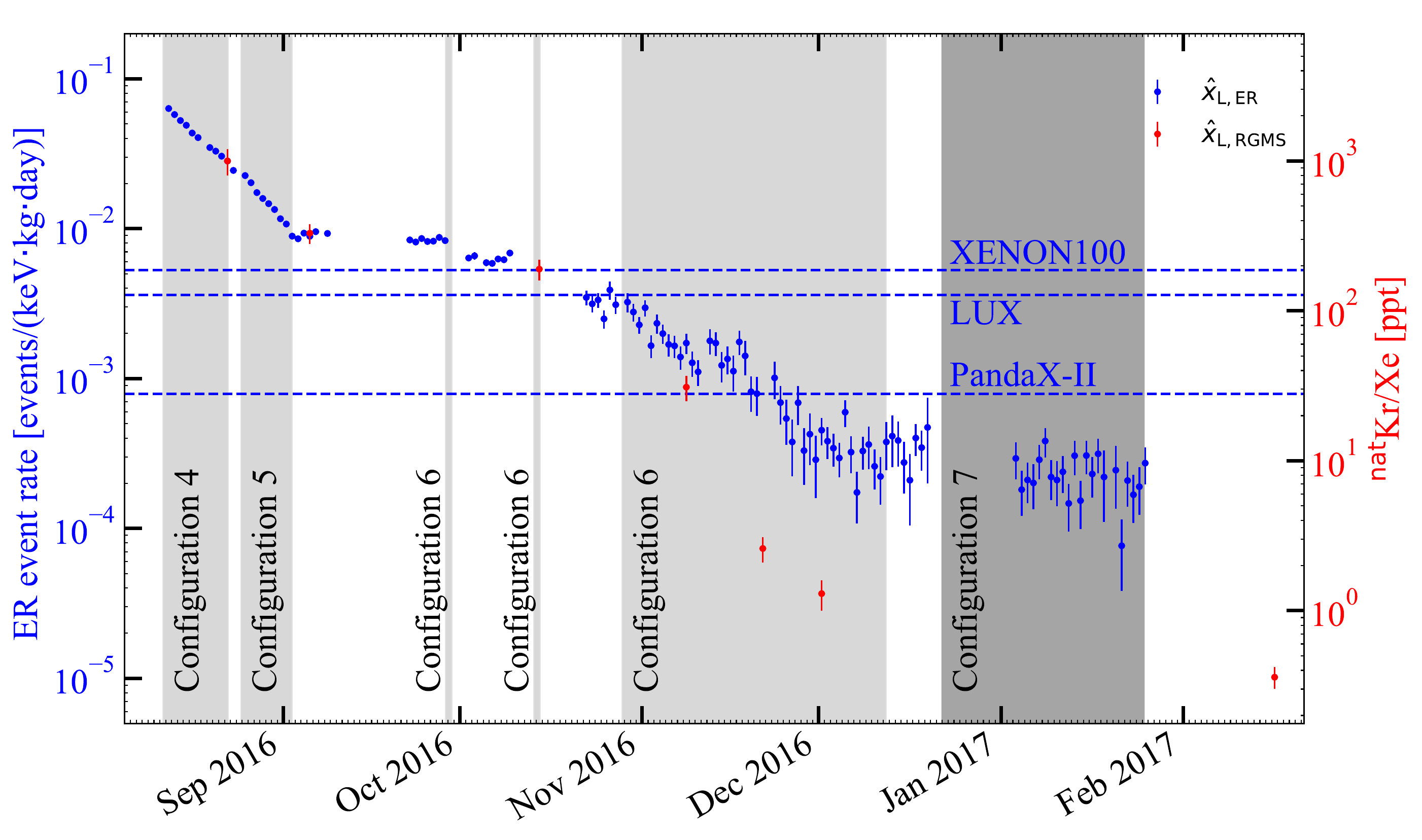}
\caption{ER event rate during online distillation campaigns in XENON1T: The ER event rate $\hat{x}_\mathrm{L,ER}$ (blue) between August 2016 and February 2017 was reduced by online krypton distillation campaigns (light grey shaded area) using \textit{C4 (Standard purification with distillation)}, \textit{C5 (Purification and distillation of gas volume alone)}, and \textit{C6 (Standard purification with upgraded gas-volume-only distillation)}. The ER event rate was further reduced by an online radon distillation campaign (dark grey shaded area) by applying \textit{C7 (Standard purification with upgraded gas-volume-only radon distillation)}. Absolute krypton-in-xenon measurements $\hat{x}_\mathrm{L,RGMS}$ with the RGMS are illustrated in red. Note that the online radon distillation had no impact on the krypton concentration. Figure based on \cite{Murra:2019:thesis}.}
\label{fig:event_rate}
\end{figure}

In figure \ref{fig:event_rate}, the resulting ER rate between August 2016 and February 2017 is visualized in blue along with the absolute krypton-in-xenon concentrations obtained from the RGMS data in red. The different online krypton distillation campaigns are shaded in light grey indicating the time periods when krypton was being removed from the system. The online radon distillation campaign, shaded in dark grey, reduced the ER rate by another \SI{20}{\percent}, but it had no impact on the krypton concentration. The details for this operation are discussed in Ref. \cite{Murra:2019:thesis}. The first RGMS data point was taken in the krypton-dominated period and was matched to the event rate. This scaling was used to convert all event rate data to equivalent krypton concentrations for the purpose of fitting the model. 
For comparison, the ER rates of other liquid xenon-based dark matter experiments such as XENON100 \cite{Aprile:2016swn}, LUX \cite{Akerib:2013tjd} and PandaX-II \cite{Cui:2017nnn} are shown. Among them, XENON1T reached the lowest background to date with the help of the online distillation method.

\textit{C4 (Standard purification with distillation)} and \textit{C5 (Purification and distillation of gas volume alone)} configurations efficiently reduced the krypton concentration inside the LXe TPC as indicated by both the event rate as well as the RGMS data. After the two short tests with \textit{C6 (Standard purification with upgraded gas-volume-only distillation)} where no large reduction can be observed, the RGMS samples still matched the event rate. During the long-term online distillation campaign with \textit{C6 (Standard purification with upgraded gas-volume-only distillation)}, the event rate starts to level off around December 2016, while the RGMS data points reveal a further absolute krypton concentration decrease. This is a clear indication that krypton is no longer the dominant ER background source. Therefore, the event rate cannot be further applied as a krypton monitoring tool starting from February 2017. The lowest krypton concentration ever documented in a xenon-based detector is \SI{360(60)}{ppq}, measured in XENON1T with the xenon sample from 16 Feb 2017. This value was sufficiently low for XENON1T, as krypton was reduced to a subdominant ER source.

In the following, the model derived in section \ref{sec:xe1t_kr_model} is fitted to the absolute krypton concentrations from the different methods mentioned above. For the different configurations, the values of $F_\mathrm{G}(t)$, $F_\mathrm{L}(t)$, $F_\mathrm{D}(t)$, $\epsilon_\mathrm{HE} = \num{0.975}$, and $F_\mathrm{con}(t)$ are derived from slow control parameters and thus are defined at all times. These variables are used as input for the model and a summary of typical values is given in table \ref{tab:model_flows}.

\begin{table}[!ht]
\caption{Typical flow values for a given configuration calculated from slow control parameters. All flows are given in [kg/d].}
\label{tab:model_flows}
\centering
\begin{tabular}{c|ccccccccc}
\hline
Name & $F_{\mathrm{G}}$ & $F_{\mathrm{L}}$ & $F_{\mathrm{con}}$ & $F_{\mathrm{D}}$ & $F_{\mathrm{1}}$ & $F_{\mathrm{2}}$ & $F_{\mathrm{3}}$ & $F_{\mathrm{4}}$ & $F_{\mathrm{5}}$ \\ \hline
C1   & 32              & 372              & 105                & 0               & 404             & 404              & 32              & 9              & 363               \\
C2   & 0              & 407              & 105                & 0               & 407             & 407              & 16              & 0              & 391               \\
C3   & 0              & 0              & 100                & 0               & 0             & 0              & 0              & 0              & 0               \\
C4   & 32              & 323              & 106                & 61               & 355             & 355              & 32              & 8              & 315                \\
C5   & 186              & 0              & 109                & 61               & 186             & 186              & 186              & 0              & 0                \\
C6   & 32              & 416              & 110                & 32               & 416             & 448              & 32              & 10              & 406                \\
C7   & 32              & 405              & 109                & 32               & 405             & 437              & 32              & 10              & 395                 \\ \hline
\end{tabular}
\end{table}

After the re-condensation by the PTR, an unkown amount of xenon evaporates while travelling through the cryopipe before reaching the LXe volume in the cryostat. Thus, the obtained values for $F_\mathrm{con}(t)$ represent only an upper limit of the condensed flow; and a free fit parameter $0 < \epsilon_{\mathrm{con}} \leq 1$ was defined to fit the data with a scaled-down condensed flow $\left(\epsilon_{\mathrm{con}}  F_\mathrm{con}(t)\right)$. This also affects the size of the evaporation flow $F_\mathrm{vap}(t)$ that is calculated from $F_\mathrm{con}(t)$. The relative volatility is fixed to  $\alpha = \num{10.5}$ (at \SI{-98}{\celsius } \cite{NIST}) in all parts of the model, except for the HE. The enrichment in the HE is unknown, and therefore the variable $\alpha_{\mathrm{HE}}$ is a free fit parameter.
The masses in both volumes are initially fixed to $M_{\mathrm{G}} = \SI{21.5}{kg}$ and $M_{\mathrm{L}} = \SI{3190}{kg}$, respectively. In the case of the LXe volume, $M_{\mathrm{L}}$ subsequently decreases during online krypton distillation campaigns due to the offgas flow $\left(F_{\mathrm{D}} - F^{\prime}_{\mathrm{D}}\right)$ that is derived from slow control. Given the large separation factor with verified outlet concentrations below \SI{48}{ppq} \cite{Aprile:2016xhi}, it is assumed that no krypton leaves the distillation system's purified outlet ($x_{\mathrm{D}} = \num{0}$). 
One unknown parameter is the migration flow $F_{\mathrm{mig}}$ accounting for the migration effect in a static LXe volume with GXe phase above, away from equilibrium. That is fitted as well. Another free fit parameter is the krypton source term $k_{\mathrm{S}}$ to account for the krypton increase in the system after the last online distillation campaign.
Furthermore, a constant background $c_{\mathrm{bg}}$ is introduced as a free parameter to take the flattening of the ER rate during the long-term operation in \textit{C6 (Standard purification with upgraded gas-volume-only distillation)} into account. With that, the ER data points are fitted to 
\begin{equation}
x_{\mathrm{L,ER}}(t) = x_{\mathrm{L}}(t) + c_{\mathrm{bg}}.
\end{equation}
All fit parameters are assumed to be time independent.

At $t = 0$, the detector is assumed to be in krypton particle equilibrium during \textit{C1 (Standard purification without distillation)}. Consequently, the change of krypton in the GXe as well as LXe is $\left(dx_{\mathrm{G}}/dt\right)|_{t=0} = \left(dx_{\mathrm{L}}/dt\right)|_{t=0} = 0$. The starting concentration in the liquid $x_{\mathrm{L,0}}$ is a free fit parameter. By solving equation (\ref{eq:kr_online_model_gxe_standard_circulation}), the corresponding krypton concentration $x_{\mathrm{G,0}}$ in the GXe can be calculated \cite{Murra:2019:thesis}. 

\begin{figure}[!!!ht]
  \centering
  \begin{subfigure}{0.76\textwidth}
    \includegraphics[width=\textwidth]{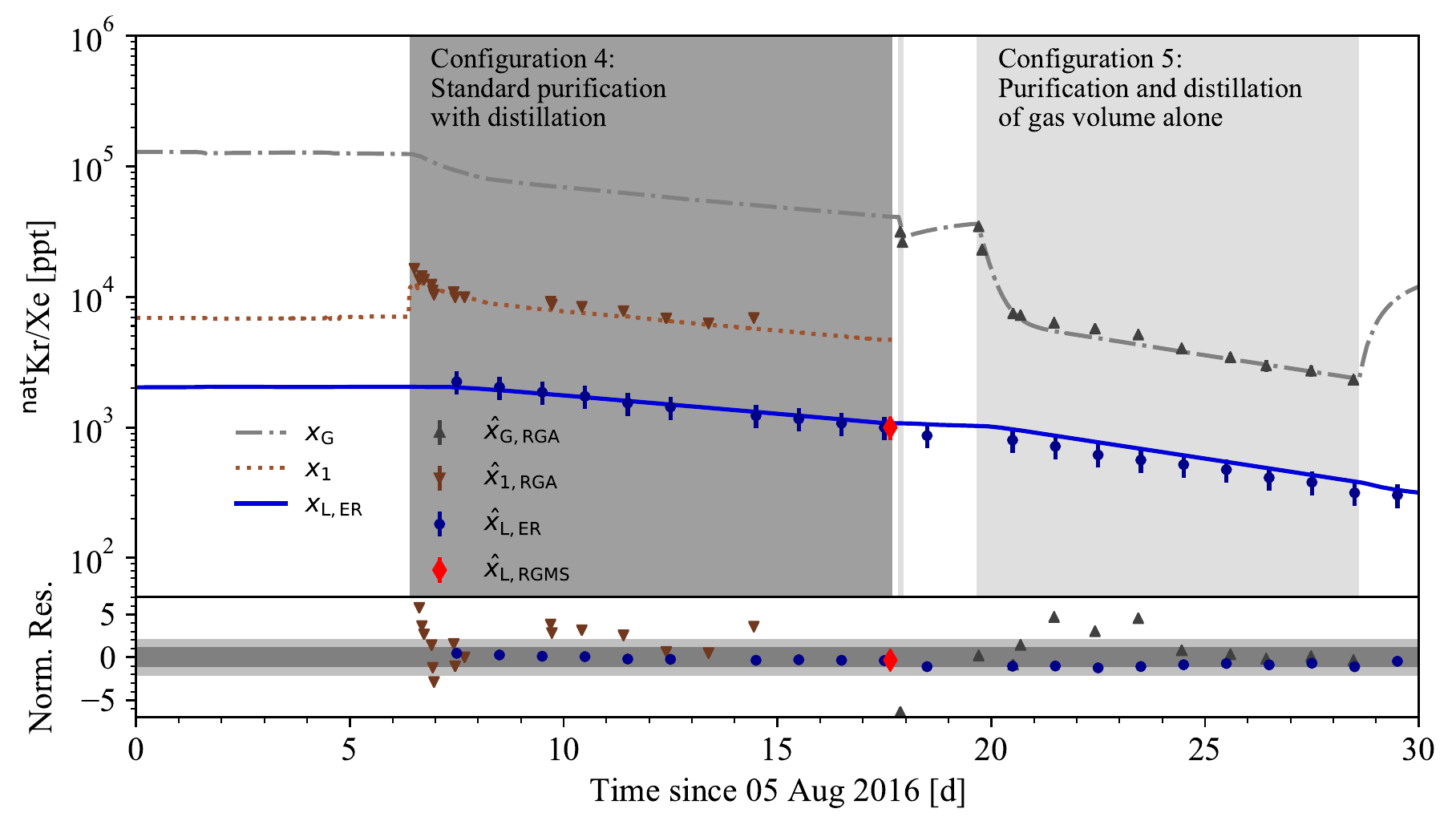}
  \end{subfigure}
   \begin{subfigure}{0.76\textwidth}
    \includegraphics[width=\textwidth]{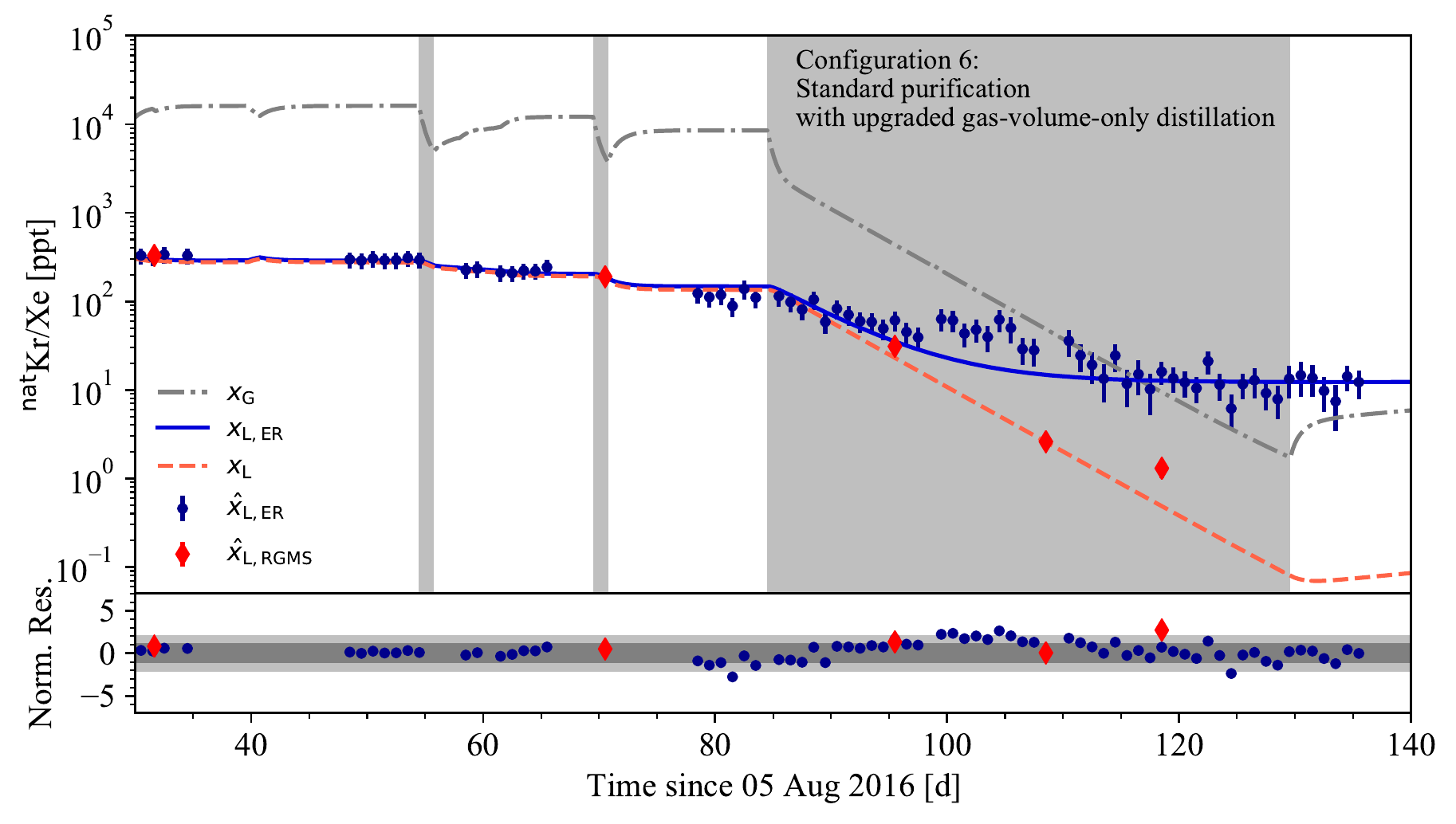}
  \end{subfigure}
  \begin{subfigure}{0.76\textwidth}
    \includegraphics[width=\textwidth]{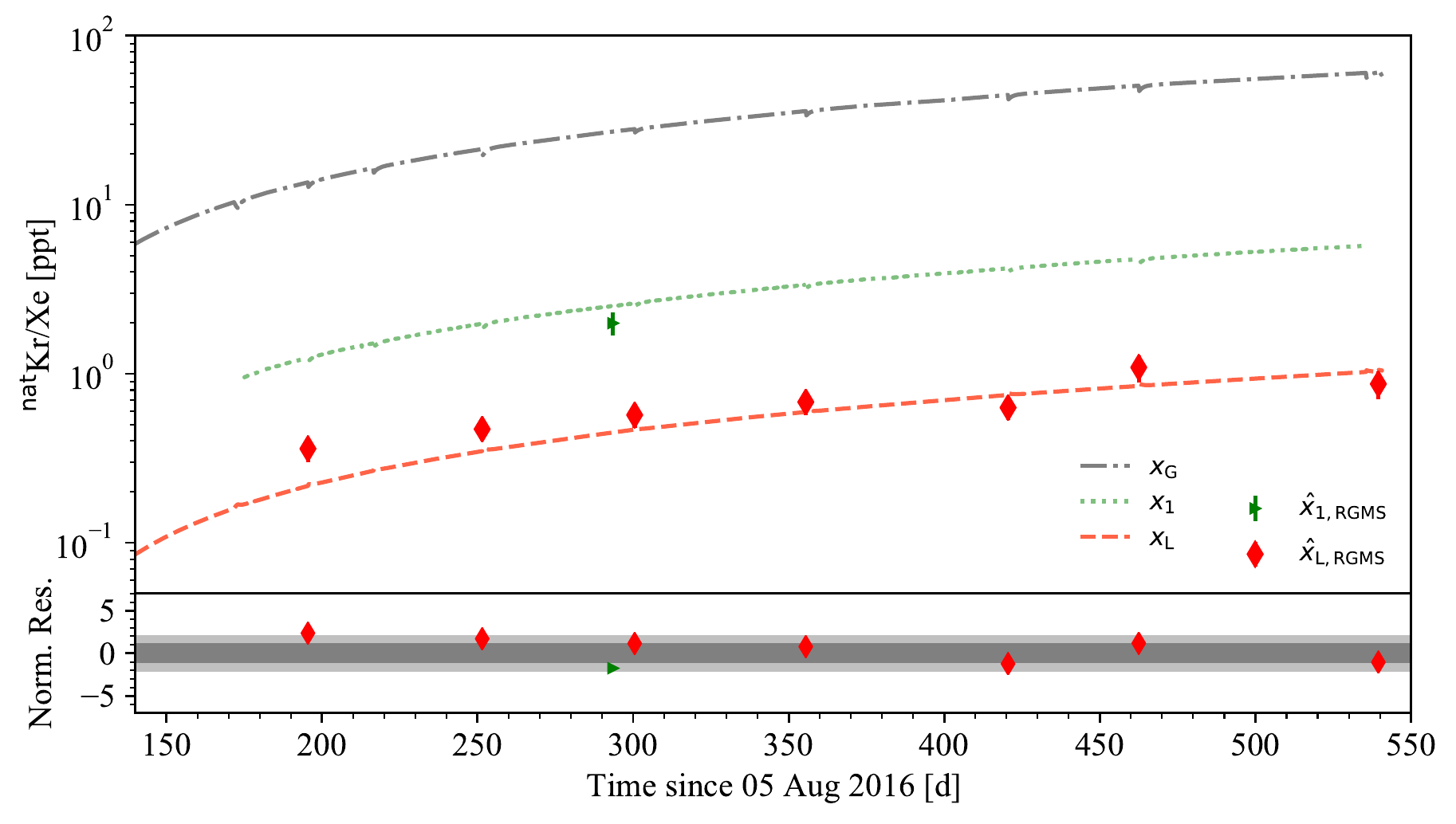}
  \end{subfigure}
\caption{Fit online distillation model to krypton data: The fit curves are labeled as $x_\mathrm{i}$, while the different measurements are denoted by $\hat{x}_\mathrm{i}$. The fit results are presented in three panels. Normalized residuals are shown in the bottom of each panel. Grey bands visualize the 1$\sigma$ and 2$\sigma$ deviation.}
\label{fig:fits_model}
\end{figure}

In order to infer the set of parameters which describes the data best, we construct for each quantity we measure ($\hat{x}_\mathrm{G,RGA}$, $\hat{x}_\mathrm{L,RGMS}$, $\hat{x}_\mathrm{L,ER}$, $\hat{x}_\mathrm{1,RGA}$, $\hat{x}_\mathrm{1,RGMS}$) a likelihood term which we combine into a final likelihood used for inference. This means that the fitting routine minimizes the model prediction with respect to the data points. For example, $x_{\mathrm{G}}$ is compared with $\hat{x}_\mathrm{G,RGA}$, $x_{\mathrm{L}}$ with $\hat{x}_\mathrm{L,RGMS}$, $x_{\mathrm{L,ER}}$ with $\hat{x}_\mathrm{L,ER}$, and $x_{\mathrm{1}}$ with $\hat{x}_\mathrm{1,RGA}$ as well as $\hat{x}_\mathrm{1,RGMS}$. The optimization is implemented as a $\chi^{2}$-minimization with iMinuit \cite{iminuit,James:1975dr} to obtain the set of best-fit parameters shown in table \ref{tab:xe1t_fit_results_online_krypton}. The corresponding model and data points are depicted in figure \ref{fig:fits_model}. For better overview, the results are divided into different time intervals including the relevant data and fit curves for each interval. The normalized residuals are shown below each plot.
%
%
%
%
%
%
%
\begin{table}[!ht]
\caption{Fit results for the online krypton distillation model.}
\label{tab:xe1t_fit_results_online_krypton}
\centering
\begin{tabular}{|c||c|}
\hline
Parameter & Result\\ 
\hline
$x_{\mathrm{L,0}}$              & \SI{2016(20)}{ppt} \\
$F_{\mathrm{mig}}$              & \SI{8.9(3)}{kg/d}			 \\
$\epsilon_{\mathrm{con}}$       & \num{0.31(1)}           \\
$c_{\mathrm{bg}}$                & \SI{12.2(9)}{ppt}          \\
$k_{\mathrm{S}}$           & \SI{10.4(7)e-12}{kg/d}             \\
$\alpha_{\mathrm{HE}}$           & $>$\num{1.45e4} (\SI{90}{\percent} C.L.)     \\  
\hline
$\chi^2$ / NDF                  & 605 / 133                     \\
\hline
\end{tabular}
\end{table}

The initial krypton concentration $x_{\mathrm{L,0}}$ in the LXe volume allows to calculate the related concentration within the GXe volume to be $x_{\mathrm{G,0}} = \SI{1.3e5}{ppt}$. The ratio between both phases yields an enhancement in the GXe volume by a factor of \num{64}, about \num{6} times larger than the relative volatility $\alpha = \num{10.5}$. As shown in equations (\ref{eq:C3_lowfactor}) and (\ref{eq:C3_highfactor}), an enhancement factor between $\alpha$ (single-stage distillation) and $\alpha^2$ (two-stage distillation) is expected in the GXe volume.

The fitted factor $\epsilon_{\mathrm{con}}$ to scale $F_{\mathrm{con}}$, and by that also $F_{\mathrm{vap}}$, is correlated with $F_{\mathrm{mig}}$. Both, $\epsilon_{\mathrm{con}}$ and $F_{\mathrm{mig}}$, contribute to terms in equations (\ref{eq:kr_online_model_gxe_standard_circulation}) and (\ref{eq:kr_online_model_lxe_standard_circulation}) that allow the krypton to move into the GXe volume. Our data does not allow to differentiate the two processes as explained in section \ref{sec:xe1t_kr_model}. However, from the best fit result, the migration flow of $F_{\mathrm{mig}}$ is the lowest with respect to the other flows used in the model. As discussed in more details in Ref. \cite{Murra:2019:thesis}, the evaporation flow $F_{\mathrm{vap}}$ and the extraction flow $F_{\mathrm{L}}$ seem to be the main drivers of the krypton removal from the LXe volume.
The background $c_{\mathrm{bg}}$ in units of a krypton concentration can be interpreted as the radon contribution to the ER rate in XENON1T and can be converted to a \isotope[222]{Rn} activity concentration of \SI{13.5(26)}{\upmu Bq/kg} \cite{Murra:2019:thesis}. This is in agreement with the value of \SI{13.6(9)}{\upmu Bq/kg} in XENON1T during December 2016 from independent in-situ $\upalpha-$spectroscopy \cite{Aprile:2020vmn}.
The source term $k_{\mathrm{S}}$ can be attributed either to external leaks or to desorption from internal detector materials. The obtained value would correspond to an air leak rate of  $k_{\mathrm{air}} = \SI{2.9(2)e-5}{(mbar \times l)/s}$, assuming a krypton air fraction of \SI{1.14e-6}{vol/vol} \cite{airliquid2018}. As the global XENON1T system was leak-checked to be below \SI{e-8}{(mbar \times l)/s}, the rate seems to be too high to originate from external leaks. The amount of residual air trapped in PTFE detector components is inferred in Ref. \cite{Aprile:PUR} from the rate of oxygen desorbing from their surfaces. The desorption rate is derived from a time-series fit to the electron lifetime, which is inversely proportional to the concentration of oxygen in the LXe volume. The result is compatible with the source term $k_{\mathrm{S}}$ found in this work, assuming a krypton air fraction as stated above, implying that desorption of krypton from PTFE is a measurable background source. This finding shows that the pumping time prior to xenon filling crucially impacts the krypton-in-xenon concentration after the krypton removal. It further shows that relying on the conventional offline krypton removal technqiues is risky. This risk can be fully avoided by applying the online distillation.

The enrichment factor $\alpha_{\mathrm{HE}}$ in the HE is constrained to be much larger than the relative volatility $\alpha$, meaning that basically all krypton entering the HE returns to the GXe volume of the cryostat. This is mainly due to the \textit{C4 (Standard purification with distillation)} data, where measurements from the GXe and LXe volumes are available, whereas the HE is not operational during \textit{C5 (Purification and distillation of gas volume alone)}. The large $\alpha_{\mathrm{HE}}$ could be due to un-modeled systematics in these measurements, or a true enhancement in the GXe solute concentration in the HE. Since the xenon flow in the HE is unidirectional, in contrast to the other volumes, the high krypton vapor pressure may make it difficult for krypton particles to enter the LXe from the GXe, leading to an enhancement as the xenon repeatedly condenses and liquifies along its path back to the cryostat, especially across the large surface in the tube-in-tube heat exchanger section.

Figure \ref{fig:fits_model} (Top) shows the time period from $t = \SI{0}{d}$ (05 Aug 2016) until $t = \SI{30}{d}$ including \textit{C4 (Standard purification with distillation)} (dark grey) as well as \textit{C5 (Purification and distillation of gas volume alone)} (light grey).

Figure \ref{fig:fits_model} (Middle) contains the time period $t = \SI{30}{d}$ until $t = \SI{140}{d}$. The two thin shaded areas correspond to the short test operations using \textit{C6 (Standard purification with upgraded gas-volume-only distillation)}, while the wide shaded area represents the long term distillation in this configuration.
According to the model, the minimum krypton concentration reached inside the LXe TPC was $x_{\mathrm{L,min}} = \SI{80}{ppq}$ at $t = \SI{137}{d}$. The corresponding concentration in the GXe volume was calculated to be $x_{\mathrm{G}}(t =  \SI{137}{d}) = \SI{5.3}{ppt}$, a factor of \num{66} larger than in the LXe volume, as also observed during \textit{C1 (Standard purification without distillation)} at the beginning of the time period investigated. Due to the source term $k_{\mathrm{S}}$, this unprecedented low concentration could not be maintained. The effective time constants $\tau_{\mathrm{eff}}$ for the exponential decrease of $x_{\mathrm{L}}$ are computed for the different online distillation configurations and are compared in table \ref{tab:online_dst_summary} along with the reduction achieved in the LXe volume for the given duration. One finds that \textit{C6 (Standard purification with upgraded gas-volume-only distillation)} is the most efficient configuration, as expected.

Figure \ref{fig:fits_model} (Bottom) illustrates the time period from $t =  \SI{140}{d}$ until $t = \SI{550}{d}$ (06 Feb 2018).
Compared to the minimum concentration, krypton increased by a factor of \num{13.5} to $x_{\mathrm{L}}(t = \SI{550}{d}) = \SI{1}{ppt}$.

Some time periods show a systematic mismatch between the model and data, indicating un-modeled effects and leading to a large ($\chi^2$/NDF) value. Given the complexity of the system and the variation of detector conditions over the time span of \SI{550}{d}, the model described the data adequately overall.

\begin{table}[!ht]
\caption{Comparison of the different online distillation configurations.}
\label{tab:online_dst_summary}
\centering
\begin{tabular}{c|ccc}
\hline
Name & Duration [d]  &  $\tau_{\mathrm{eff}}$ [d] & Reduction factor in LXe volume     \\ \hline
C4   &  11.3         &    15.3                   & 1.9          \\
C5   &  8.9         &    8.7                    & 2.7          \\
C6   & 45.1         &    6.0                    & \num{1.71e3} \\ \hline
\end{tabular}
\end{table}
\section{Argon removal}
\label{sec:argon}
A gaseous \isotope[37]{Ar} source was deployed in XENON1T in October 2018 before its decommissioning \cite{Aprile:Ar37} that allowed for a calibration down to energies of \SI{2.8}{keV} and \SI{0.27}{keV} via electron capture \cite{TabRad_v7}.

At its critical temperature of \SI{-123}{\degreeCelsius}, argon features a vapor pressure of about \SI{50}{bar} \cite{NIST}. At the LXe temperature of \SI{-96}{\degreeCelsius}, the argon vapor pressure is not defined. Therefore, the relative volatility is assumed to be $\alpha_{\mathrm{Ar}}>25$ for a xenon pressure of \SI{2}{bar}. The larger volatility with respect to krypton should make the online distillation more efficient, i.e. reducing \isotope[37]{Ar} with a faster effective time constant than krypton.

The \isotope[37]{Ar} event rate evolution in the LXe volume is fitted with the online distillation model of this work including the time periods of the injection, the calibration as well as the removal.
Configurations \textit{C1 (Standard purification without distillation)}, \textit{C2 (Evaporated-liquid-only purification without distillation)}, and \textit{C6 (Standard purification with upgraded gas-volume-only distillation)} were used during the \isotope[37]{Ar} calibration campaign. In order to apply the model to the argon data, small modifications need to be done to the coupled differential equations (\ref{eq:kr_online_model_gxe_standard_circulation}) and (\ref{eq:kr_online_model_lxe_standard_circulation}) for the different configurations.

First, the relative volatility was changed to a fit parameter with $\alpha_{\mathrm{Ar}}>25$ since argon is supercritical at LXe temperature, and the source term $k_{\mathrm{S}}$ was set to zero. Second, in contrast to \isotope[85]{Kr}, the reduction of the event rate due to the radioactive decay of \isotope[37]{Ar} needs to be taken into account. Therefore, the terms $\left(-\lambda_{\mathrm{Ar37}}  x_{\mathrm{G}}\right)$ and $\left(-\lambda_{\mathrm{Ar37}}  x_{\mathrm{L}}\right)$ with $\lambda_{\mathrm{Ar37}} = \SI{0.01980}{d^{-1}}$ are added to the equations for GXe and LXe, respectively. The parameter $\epsilon_{\mathrm{con}}$ was fixed to \num{0.31} as taken from the krypton removal fit to reduce the number of free parameters, since it is anyway strongly correlated with the migration flow $F_{\mathrm{mig}}$ and the relative volatility $\alpha_{\mathrm{Ar}}$.

The gaseous source was added to the PUR system during \textit{C1 (Standard purification without distillation)}. In total, three \isotope[37]{Ar} injections were done during the calibration campaign. For simplicity, each injection in this model is added directly into the GXe volume as a delta peak $+ k_{\mathrm{inject}}$ for its given injection time.

\begin{table}[!h]
\caption{Fit results for the online argon distillation model.}
\label{tab:xe1t_fit_results_online_argon}
\centering
\begin{tabular}{|c||c|}
\hline
Parameter & Result\\ 
\hline
$F_{\mathrm{mig}}$              & \SI{5.3(3)}{kg/d}			 \\
$\alpha_{\mathrm{Ar}}$       & \num{50.9(2)}           \\
$A_{\mathrm{inject1}}$           & \SI{2.6(1)}{Bq}             \\ 
$A_{\mathrm{inject2}}$           & \SI{4.3(1)}{Bq}  \\ 
$A_{\mathrm{inject3}}$           & \SI{2.1(1)}{Bq}             \\ 
\hline
$\chi^2$ / NDF                  & 766/513                     \\
\hline
\end{tabular}
\end{table}

\begin{figure}[!h]
\centering
\includegraphics[width=0.95\textwidth]{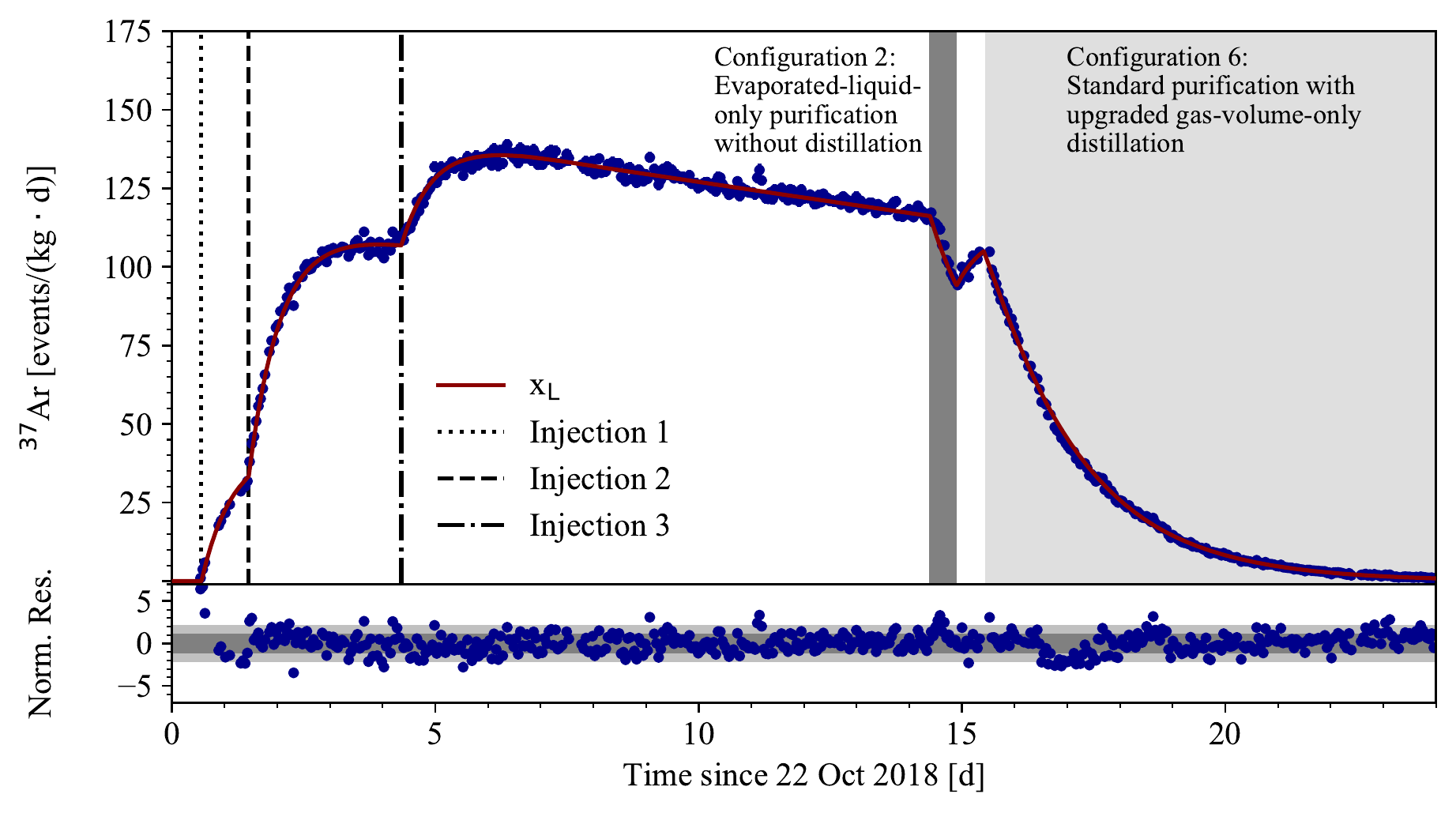}
\caption{Fit online distillation model to argon data: Three \isotope[37]{Ar} injections into the GXe volume led to an increase in \isotope[37]{Ar} events inside the LXe volume. The event evolution during \textit{C1 (Standard purification without distillation)} (white), a short period \textit{C2 (Evaporated-liquid-only purification without distillation)} (dark grey), and \textit{C6 (Standard purification with upgraded gas-volume-only distillation)} (light grey) is described by $x_{\mathrm{L}}$.}
\label{fig:ar37_evolution_fit_1}
\end{figure}

The results of the fit are summarized in table \ref{tab:xe1t_fit_results_online_argon} where the injections into the GXe volume were directly transformed into an activity $A_{\mathrm{inject}}$ for better readability. The evolution of the \isotope[37]{Ar} event rate in the LXe volume and the corresponding fit curve $x_{\mathrm{L}}$ are plotted in figure~\ref{fig:ar37_evolution_fit_1}. The corresponding evolution for the GXe volume $x_{\mathrm{G}}$ was omitted for better visualization. The ratio $x_{\mathrm{G}}/x_{\mathrm{L}}$ shows an \isotope[37]{Ar} enrichment in the GXe volume by a factor \num{100} during \textit{C1 (Standard purification without distillation)}, larger than the fitted value for $\alpha_{\mathrm{Ar}}$. This behaviour was also observed in the krypton case, and is expected from the discussions in section \ref{sec:xe1t_kr_model}.
The migration flow $F_{\mathrm{mig}}$ and the relative volatility $\alpha_{\mathrm{Ar}}$ are part of the migration term in the differential equations and thus, are correlated. Therefore, a comparison of the migration flow to the case of krypton is not possible. Furthermore, the model can effectively describe the argon evolution, but cannot be used to measure precisely the relative volatility for argon in xenon since the global system is too complex and not designed for such a measurement.

The effective time constant for the \isotope[37]{Ar} decrease during the online distillation campaign is $\tau_{\mathrm{eff,Ar}} = \SI{1.7}{d}$ and shows an efficient reduction within two weeks. This makes it possible to consider \isotope[37]{Ar} as a regular calibration source for multi-tonne xenon detectors such as XENONnT, LZ, PandaX-4T, and DARWIN.

\MM{In the case of XENONnT, the main xenon handling infrastructure relevant for the online distillation method stays the same with the addition of a LXe purification system. The GXe volume mass remains about \SI{20}{kg}, while the LXe volume mass is three times larger than in XENON1T. The external heat input is expected to be also about three times larger, and scales with the larger surface of the XENONnT cryostat. Thus, the time constant of the evaporation term, one of the main drivers of the solute transport, is expected to be similar in XENONnT. The extraction flow $F_{\mathrm{L}}$, the other main driver, remains the same, and thus, its time constant is larger resulting in a slower overall removal in XENONnT. However, the new LXe purification system extracts large flows of xenon of about \SI{2}{LPM} directly in liquid form from the cryostat. Unless the returning LXe is sub-cooled through other means, the net effect would be an additional heat input to the cryostat. This could result in an enhanced transport of lighter components from the LXe to the GXe volume by xenon evaporation yielding a boost in the removal time constant. The size of this effect cannot be estimated, but will be further investigated in XENONnT.}
\section{Conclusion}
\label{sec:conclusion}
A novel online distillation technique was developed for the XENON1T experiment to reduce more volatile intrinsic noble gases inside the LXe TPC during its normal operation. The method is based on a continuous distillation of the gaseous xenon volume of the detector with the help of the XENON1T cryogenic distillation column. 
The main focus was to lower the krypton-in-xenon concentration for the first XENON1T science run. A confirmed concentration of \SI{360(60)}{ppq} in the liquid xenon detection volume was achieved, the lowest measured in a dark matter detector to date. The online distillation was stopped as soon as krypton was a negligible background with respect to radon, but before reaching its limits.

In addition, the online distillation method was applied to reduce \isotope[37]{Ar} after it was deployed as a calibration source for low energies down to \SI{2.8}{keV}. 
Usually, the \isotope[37]{Ar} half-life of \SI{35.01}{d} is too long for regular use of this source. However, the online distillation reduced the \isotope[37]{Ar} event rate inside the LXe TPC back to a negligible level within two weeks.

An online distillation model was developed to describe several detector configurations based on coupled differential equations for the krypton-in-xenon concentrations within the detector's LXe and GXe volumes. The krypton time evolution in the system was monitored via the event rate within the LXe TPC itself, as well as via several extracted xenon samples. The model was successfully fitted to the data over a time span of 550 days, including the commissioning, science run 0 and science run 1 of XENON1T. The gained knowledge regarding the krypton transport inside the different xenon handling subsystems helps for the development of future experiments.

With small adaptions due to the properties of argon, such as a higher volatility compared to krypton, the online distillation model was validated by successfully fitting the \isotope[37]{Ar} induced event rate evolution in the LXe volume.

Whenever an online distillation is performed, a small offgas flow needs to be removed from the distillation system, and thus from the global system. This xenon loss needs to be balanced by over-filling (re-filling) the detector directly via the distillation column before (after) the operation. In future applications, this process can be optimized and automated to keep the xenon mass in the global system constant by supplying additional xenon from the storage system to the DST system's inlet. \MM{Other influences on the LXe TPC performance were not observed.}

In summary, the online distillation method can be applied at any given time during the lifetime of an experiment, either to reduce impurities after initial detector filling or after accidental leaks due to handling errors or hardware failures, or to remove more volatile noble gas calibration sources. Since the concept was proven for argon and krypton, also helium and neon should be efficiently removed.

\section*{Acknowledgment}
We gratefully acknowledge support from the National Science Foundation, Swiss National Science Foundation, German Ministry for Education and Research, Max Planck Gesellschaft, Deutsche Forschungsgemeinschaft, Helmholtz Association, Dutch Research Council (NWO), Weizmann Institute of Science, Israeli Science Foundation, Fundacao para a Ciencia e a Tecnologia, R\'egion des Pays de la Loire, Knut and Alice Wallenberg Foundation, Kavli Foundation, JSPS Kakenhi in Japan and Istituto Nazionale di Fisica Nucleare. This project has received funding/support from the European Union’s Horizon 2020 research and innovation programme under the Marie Sk\l odowska-Curie grant agreement No 860881-HIDDeN. Data processing is performed using infrastructures from the Open Science Grid, the European Grid Initiative and the Dutch national e-infrastructure with the support of SURF Cooperative. We are grateful to Laboratori Nazionali del Gran Sasso for hosting and supporting the XENON project.%

\let\doi\relax

\bibliographystyle{ptephy}
\bibliography{sample}

%
\end{document}